\documentclass[showpacs,preprintnumbers,
 amsmath,amssymb,nofootinbib,superscriptaddress]
{revtex4}
\usepackage{graphicx}
\usepackage{epsfig}
\usepackage{bm}
\usepackage{amsfonts}

\newcommand{\eq}{\begin{eqnarray}}
\newcommand{\eqx}{\end{eqnarray}}
\newcommand{\ba}{\begin{equation}}
\newcommand{\ea}{\end{equation}}
\newcommand{\f}[2]{\frac{#1}{#2}}

\newcommand{\cor}[1]{\left\langle{#1}\right\rangle}

\newcommand{\bit}{\begin{itemize}}
\newcommand{\eit}{\end{itemize}}

\newcommand{\vp}{\varphi}
\newcommand{\xb}{x_\bot}
\newcommand{\vb}{v_\bot}

\newcommand{\bR}{\bar R}

\def\la{\label}
\def\nn{\nonumber \\}
\def\bi{\bibitem}

\def\lam{\lambda}

\def\d{\partial}

\def\al{\alpha}

\def\b{\beta}
\def\ve{\varepsilon}
\def\ub{u_{\bot}}

\begin{document}

\title{On an exact hydrodynamic solution for the  elliptic flow}

\author{Robi Peschanski }
\email{robi.peschanski@cea.fr} \affiliation{Institut de Physique
Th\'eorique, CEA, IPhT, F-91191 Gif-sur-Yvette, France, CNRS, URA
2306 }

\author{Emmanuel N.~Saridakis }
\email{msaridak@phys.uoa.gr} \affiliation{Department of Physics,
University of Athens, GR-15771 Athens, Greece}

\begin{abstract}
Looking for the underlying hydrodynamic mechanisms determining the
elliptic flow we show that for an expanding relativistic perfect
fluid the transverse flow may derive  from a solvable hydrodynamic
potential, if the entropy is transversally  conserved and the
corresponding expansion ``quasi-stationary'', that is mainly
governed by the temperature cooling. Exact solutions for the
velocity flow coefficients v$_{2}$ and the temperature dependence
of the spatial and momentum anisotropy are  obtained and shown to
be in agreement with the elliptic flow features of heavy-ion
collisions.
\end{abstract}

 \maketitle

\section{Introduction}
The hydrodynamic description of the formation and development of a
quark-gluon plasma (QGP) in high-energy  heavy-ion collisions has
met a considerable success \cite{Hirano}. In particular, the
hydrodynamic features seem to be, at least partly, required in
order to take into account   the second Fourier coefficient $v_2$
of the  transverse flow of particles, the so-called  {\it elliptic
flow}. One writes \cite{Wong,Ollitrault:1992bk} for the azimuthal
multiplicity distribution \eq \f {dN}{d\vp}= \f N{2\pi}\left\{1+
2v_2\ cos(2\vp)+\cdots\right\} \la{azimuth}, \eqx discarding for
simplicity other Fourier coefficients non-relevant here. The
experimentally observed  values of $v_2$, which are due to the
anisotropy of the initial state collisions at nonzero impact
parameter, are sizeable enough to require important collective
effects of particle production. These  are better reproduced by
hydrodynamical properties of the flow in some early stage of the
quark-gluon plasma formation.

The theoretical estimates of the elliptic flow are  obtained from
numerical studies based on various versions of the hydrodynamic
models. Indeed, a full study requires not only to deal with the
solution of the relativistic hydrodynamic equations but also with
the definition of appropriate initial conditions and  a model for
the mutation of the QGP pieces of fluid into particles. The
numerical studies ($cf.$ \cite{Hirano}) reveal that the QGP as
a fluid is ``almost perfect'' since its viscosity is remarkably
weak, even if the model dependence may account for some variation
on the quantitative estimates. This observation has a considerable
theoretical impact, since it points to a strongly coupled plasma,
guiding a large part of  theoretical interest towards strongly
coupled gauge field theories.

Our goal in the present work is to try and identify by explicit
analytic solutions the basic hydrodynamic mechanisms at work in
the elliptic flow. For this sake, we have to simplify (or even
idealize) the  description of the QGP formation in a heavy-ion
collision while keeping the main physical ingredients. Among other
simplifications that we will discuss now, our study will assume
the QGP to be a perfect fluid without viscosity. We will also
restrict our analysis to the transverse flow in the central
rapidity region where the hydrodynamic description is better
suited.

The main characteristic  feature of the hydrodynamic description
of QGP formation in heavy-ion collisions appears to be a
nontrivial  combination of: a) the  large longitudinal momentum
and energy boosts provided to the created medium by the initial
state, b) the (presumably fast) equilibration of the energy
density and  all three pressure components due to local
thermalization required by hydrodynamics. As a matter of fact, the
first stage of  the hydrodynamical description of particle
production in high energy collisions is mainly governed by the
longitudinal flow, that is the expansion of the relativistic fluid
in (1+1) dimensions. Already the pioneering papers of the
hydrodynamic approach \cite{landau,bjorken} based their analysis
on this property.

In the mean time, the 4-dimensional hydrodynamical feature of the
system is kept with the thermodynamic relations which, through
the local temperature $T$ and the Equation of State (EoS), lead
to \eq \la{temperature} \f T{T_0}\sim
\left(\f{\tau_0}\tau\right)^{c_s^2}\quad;\quad\quad \tau
\equiv\sqrt{x_0^2+x_3^2}\quad\ ;\quad\quad
 c_s^2=\f{dp}{de}=\f{sdT}{Tds}\ ,
\eqx where $\tau $ is the proper-time,  $c_s$ the speed-of-sound,
which we will assumed to be constant in the following, and $e,p$
the energy and (isotropic) pressure density, respectively. Our
remaining notations are  $s$  for the entropy density,
$x_{\mu=\{0,\cdots 3\}}$ for the space-time coordinates
($x_0\equiv t$), and $u_{\mu=\{0,\cdots 3\}}$ for the fluid
4-velocity (with lower indices) in the Minkowski metric
$\eta^{\mu\nu}$, with signature $(1,-1,-1,-1)$ satisfying the
normalization condition
\begin{equation}
 u_\mu u^\mu\equiv u_0^2-u_3^2-u_\bot^2=1\ ;\quad u_\bot^2\equiv u_1^2+u_2^2\ .
\la{norm}
\end{equation}

Our idea for studying the transverse motion of the fluid is that
it is also driven  by the longitudinal evolution, but slowly
enough to be considered as ``quasi-stationary'', that is in such a
way that its time evolution is essentially related to the
temperature cooling. Indeed, the seed of transverse  momenta is
indirect and there should be, at least during some first stage of
the hydrodynamic evolution, no strong back-reaction on the
longitudinal motion. The ``quasi-stationarity'' hypothesis will
allow for an exact solution for the elliptic flow and in general
for the hydrodynamic  regime in the transverse plane. To be
 concrete we state the following  conjectured properties of the hydrodynamic flow:\\
\begin{itemize}
\item {\bf a)} $Tranversally \ isentropic.$ Since the overall entropy
should be conserved, we will conjecture that the transverse flow
is itself (approximately) isentropic, $i.e.$ we write the following equation
\begin{equation}
\left[\partial_{x_1}(s u_1)+
\partial_{x_2}(s u_2)\right]_{transverse}=0\ .
 \label{transentropy}
\end{equation}

\item {\bf b)} $Quasi-stationary:$ Since the time dependence of the
transverse entropy distribution is absent from
\eqref{transentropy}, we will close the equations for the
transverse flow by assuming that its hydrodynamic evolution  is
smooth enough to be driven only by the temperature change.  We
thus consider the equation and solutions of a
temperature-dependent stationary flow, with a source emitting a
fluid at a given temperature (and thus transverse speed, see further).
\end{itemize}

Hence, in the regime when   the longitudinal expansion is
dominant, the transverse motion is conjectured to be smoothly
driven by the overall local temperature of the fluid, which
provides  the 4-dimensional feature\footnote{In that respect, our
picture is different from a purely transverse hydrodynamic flow
\cite{bialas}.} of the system through the thermodynamic relations
\eqref{temperature}.

The plan of the paper is the following: in the next section II,
using the hydrodynamic potential for a stationary flow \cite{KK},
we derive the analytic equation obeyed by the azimuthal
distribution of entropy and thus the elliptic flow. Then, in
section III, we find the exact solutions for the transverse flow. In
section IV we apply our solution showing that the obtained elliptic
flow retain good qualitative features observed in reality or in
realistic numerical hydrodynamic model studies.  A discussion of
the hypotheses and our conclusions and outlook form the final
section V.

\section{Azimuthal entropy distribution}

As we shall see now, the conditions {\bf a)} and {\bf b)} lead to  nontrivial
properties of  the fluid and to analytic solutions for  the
elliptic flow. It is obtained from an  hydrodynamic (KK) potential
\cite{KK} derived by  Khalatnikov and Kamenshchik for a stationary
transverse isentropic flow. The ``quasi-stationary'' hypothesis
allows us to extend its applicability to a slow transverse motion
of the fluid and to find the general exact solution of the
elliptic flow. In fact,  the existence of a  hydrodynamical
potential  obeying a linear equation is known since long
\cite{khal,bi} for the longitudinal evolution. Recently \cite{us},
it was possible to express interesting analytic solutions for the
entropy distribution $dS/dy$ where the (hydrodynamic) rapidity is
defined by $y={\scriptstyle \f 12} \log (u_0\!+\!u_3)/\log
(u_0\!-\!u_3)$. We shall follow the same method as in \cite{us}
for the transverse flow case, and find the general solution of the
KK potential in order to obtain the azimuthal distribution of the
entropy $dS/d\vp$ giving access to the elliptic flow.

However, one crucial difference of the transverse w.r.t. the
longitudinal case is the velocity-temperature relation between
$u_0$, the time component of the velocity (and thus also
$u_\bot=\sqrt{u_0^2\! -\!1}$, the modulus of the transverse one)
to the local temperature. This comes from the  relativistic
Bernoulli relation, verified by a stationary fluid \cite{Landau},
namely
\begin{equation}
Tu_0= T \sqrt{1+u_\bot^2}= T_0 \ .
\label{Bernouilli}
 \end{equation}
In the same conditions, the whole evolution  from some initial
temperature to the final freeze-out one is constrained to be in
the  {\it supersonic} regime \cite{Landau}, as we will verify
later through our equations. The condition writes
 \begin{equation}
v_\bot\equiv \f\ub{u_0}=\f\ub{\sqrt{1+\ub^2}}= \left\{1-\left(\f
T{T_0}\right)^2\right\}^{1/2} \ge \
c_s\ .
 \label{sound}
\end{equation}
Hence, the isentropic transverse evolution starts at a given
temperature $T_I$ such that
 \begin{equation}
 {T_I}\le T_s\equiv {T_0}\ \sqrt{1-c_s^2},
 \label{Tsound}
\end{equation}
and the velocity increases when the temperature decreases from
$T_I$, reaching eventually ultra-relativistic values before
hadronization. For convenience, we will from now on  introduce the
variable
\begin{equation}
l={\scriptstyle \f 12} \log\left[1-\left(\f T {T_0}\right)^2\right]
= {\scriptstyle \f 12} \log\left[\f {u_\bot^2}{1+u_\bot^2}\right]=\log \vb
\label{l}.
\end{equation}

The derivation of the KK potential comes briefly as follows
\cite{KK}: Together with the transverse entropy conservation
\eqref{transentropy}, the  equations for the transverse flow close
with the projection to the transverse plane of the energy-momentum
conservation relation $\partial_\mu T_{nu^\mu}=0$, again by
neglecting the time derivatives w.r.t. the transverse gradients.
After nontrivial transforms, presented in  appendix
\ref{appstatflow}, one obtains the system of equations \eq
\!\partial_{x_1}(s u_1)+\
\partial_{x_2}(s u_2)&=&0 \nn
 \partial_{x_1}(T u_2)-\partial_{x_2}(T u_1)&=&0
 \label{derivative}.
  \eqx
Then, using  the ``hodograph'' \cite{khal,bi,KK,Landau,us}
inversion of variables $(x_1,x_2)\to (l,\vp)$ and combining
Eqs.\eqref{transentropy} and \eqref{derivative}, one arrives  at
the formulae expressing  the kinematic (now dynamical) variables
$(x_1,x_2)$ in terms of the hydrodynamic variables through
a suitably defined  KK potential function
$\chi(\varphi,l),$ namely
\eq
\xb(\vp,l)&=&\frac{e^{-l}}{T_0}\sqrt{\left(\frac{\partial
\chi}{\partial l}\right)^2+\left(\frac{\partial \chi}{\partial
\vp}\right)^2}\nn
\alpha(\vp,l)&=&\vp+\arctan\left[{\frac{\partial\chi}{\partial\vp}}
\left({\frac{\partial\chi}{\partial
l}}\right)^{-1}\right]\ ,
\label{alpha}
\eqx
where we have parameterized
\begin{eqnarray}
u_1&=&u_\bot \cos \vp\quad \quad u_2=u_\bot \sin \vp\nonumber \\
x_1&=&x_\bot \cos \al \quad \quad x_2=x_\bot \sin \al\ .
\label{ansatz}
\end{eqnarray}

The KK potential function $\chi(\varphi,l)$ is solution of
 a linear equation obtained by closing the hydrodynamic equations
 system  using
the EoS
\begin{equation}
\left(1-\f{e^{2l}}{c_s^2}\right)\ \f{\partial^2\chi}{\partial
\vp^2}+\left(1-e^{2l}\right)\ \frac{\partial^2\chi}{\partial
l^2}+\left(1\!-\f{1}{c_s^2}\right)\ e^{2l}\ \frac{\partial
\chi}{\partial l}=0\
 \label{KK2}.
\end{equation}
Note the zero coefficient at $e^{l}=c_s$
which signals the supersonic bound (\ref{sound}, \ref{Tsound})
at $T=T_s.$ In fact the system expands in the vacuum for $T \le T_s,$
while it is compressed when $T > T_s,\ cf.$ \cite{landau}. Hence
the physical solutions are restricted to the
supersonic range $T \le T_I\le T_s.$

The KK potential and its equation have been reproduced from
\cite{KK}. The calculation of the entropy distribution is now
parallel to the one \cite{us} (see \cite{milekhin} for an early version)
used in the (1+1)
dimensional case. Considering an entropy flux normal to the
tangential line element $(dx_1,dx_2),$ one has to compute
\begin{equation}
dS=su_2dx_1-su_1dx_2\ .
 \label{milekh}
\end{equation}
Using the formulae (\ref{alpha}) for the expression of the line element
in terms of the potential,
one has
\begin{eqnarray}
-\frac{e^{-l}}{T_0}\frac{\partial\chi}{\partial l}&=&x_1\cos\vp+x_2\sin\vp=
\xb\cos(\alpha-\vp)
\nonumber\\
-\frac{e^{-l}}{T_0}\frac{\partial\chi}{\partial
\vp}&=&x_2\cos\vp-x_1\sin\vp=\xb\sin(\alpha-\vp)
 \label{hodograph},
\end{eqnarray}
which, by differentiation with respect to $l$ and $\phi,$ gives
\begin{equation}
dS=\ \f {sT_0}T\ \left\{\left[\f {\partial^2 \chi}{\partial
l\partial \vp}- \f {\partial \chi} {\partial \vp}\right]
dl+\left[\f {\partial^2 \chi}{\partial \vp^2}+ \f {\partial \chi}
{\partial l}\right] d\vp\right\},
 \label{entropydistrib}
\end{equation}
where we used the relation $u_\bot e^{-l}\equiv u_0= {T_0}/T.$

At fixed temperature (and thus fixed $l$), which is the case considered further on,
  one gets the azimuthal entropy distribution
\begin{equation}
\frac{dS}{d\vp} = \ \f {sT_0}T\ \left[\frac{\partial^2
\chi(\vp,l)}{\partial\vp^2}+\frac{\partial\chi(\vp,l)}{\partial
l}\right]=\ \f {sT}{T_0(1-e^{2l}/c_s^{2})}\
\left[\frac{\partial\chi(\vp,l)}{\partial
l}-\frac{\partial^2 \chi(\vp,l)}{\partial
l^2}\right]\ ,
\label{dentropydphi}
\end{equation}
where   the second expression comes from the KK potential equation \eqref {KK2}.
 Note again the singular coefficient at $T_s,$ corresponding to the lower bound
 of temperature.

\section{Exact solution of the transverse flow}
\label{ellflow}

In  our idealized hydrodynamic framework, without hadronization,
one relates the entropy distribution to  multiplicity, $dS/S \sim
dN/N.$ Hence, the  elliptic flow is defined by the azimuthal
entropy distribution \eqref{dentropydphi} through a Fourier
expansion similar to \eqref{azimuth}, namely
 \eq
\f {dS}{d\vp}= \f S{2\pi}\left\{1+ 2v_2\ cos(2\vp)+\cdots\right\}\ ,
\la{azimuthS}
\eqx
and thus
\begin{equation}
v_2=\frac{\int
d\vp\,\cos(2\vp)\,\frac{dS}{d\vp}(\vp)} {\int
d\vp\,\frac{dS^{}}{d\vp}(\vp)}\ .
\label{v2}
\end{equation}

The  eccentricity  can be  obtained  as a function of temperature
(or $l$) through Eqs.\eqref{hodograph} in terms of the KK
potential $\chi(\vp,l)$ as:
\begin{equation}
\varepsilon \equiv\frac{\langle
x_2^2-x_1^2\rangle}{\langle
x_2^2+x_1^2\rangle} \equiv \frac{\int
d\vp\left(x_2^2(\vp,l)\!-\!x_1^2(\vp,l)\right)}{\int
d\vp\left(x_2^2(\vp,l)\!+\!x_1^2(\vp,l)\right)}= \frac{\int
d\vp\left\{ \cos2\vp\left[
\left(\frac{\partial\chi}{\partial\vp}\right)^2-\left(\frac{\partial\chi}{\partial
l}\right)^2\right]+2\sin2\vp\ \frac{\partial\chi}{\partial\vp}\frac{\partial\chi}{\partial
l}\right\}}{\int d\vp\left[
\left(\frac{\partial\chi}{\partial\vp}\right)^2+\left(\frac{\partial\chi}{\partial
l}\right)^2\right]}\ .
\label{eccentricity}
\end{equation}
Note the  characteristic feature of the hodograph method: a
geometrical parameter, here $\ve$, is expressed in terms of
dynamical ones, here the temperature. Once the solution is found,
one has to invert these relations in order to restore the
hierarchy between ``cause'' and ``effect''.

We can now proceed by looking for the general solution resulting
from the KK potential solution of \eqref{KK2}. For this sake, it
is convenient to expand the potential
\begin{equation}
\chi(\vp,l)=\beta_0(l)+\sum_{p=1}^\infty
 \b_{p}(l)\cos(2p\,\vp)
\label{solchi}
\end{equation}
 in Fourier coefficients $\b_{p}(l)$ which verify the equations
\begin{eqnarray}
\left(e^{2l}-1\right)\b''_{p}(l)+e^{2l}(c_s^{\!
-2}-1)\b'_{p}(l)-4p^2 \left(c_s^{\!
-2}\,e^{2l}-1\right)\b_{p}(l)=0\ , \la{coeffs}
\end{eqnarray}
where primes denote derivatives with respect to $l$.

As it is well known, the solution is in general a suitable
combination, with constant coefficients, of two independent
solutions of the second-order equations \eqref{solchi}. Using
standard textbooks, one finds for $p \ne 0$
\begin{eqnarray}
\b_{p}(l)  &=&  c_{p}^{(1)}\ \b_{p}^{(1)}+c_{p}^{(2)}\
\b_{p}^{(2)}
\nonumber\\
&\equiv&  c_{p}^{(1)}\,(-)^{p+1} e^{2p\,l} \
_2F_1{\scriptstyle{\left(p+ \frac 14 \{c_s^{ -2}-1\} -
\sqrt{\{c_s^{   -2}    -  1\}^2/16  +c_s^{   -2}p^2},\ p + \frac
14 \{c_s^{   -2}    - 1\} + \sqrt{\{c_s^{ -2}    - 1\}^2/16
  +c_s^{   -2}p^2},1+2p\ ;\ {\bf e^{2l}}\right) }}
\nonumber\\
  &&+\ \  c_{p}^{(2)}\  G^{2,\,0}_{2,\,2} \left({\bf e^{2l}}\ \Big|
\begin{array}{cc}
\frac{5-c_s^{   -2}}{4}-\sqrt{\left(\frac{c_s^{
-2}-1}{4}\right)^2+c_s^{   -2}p^2}\ \ , &
\frac{5-c_s^{   -2}}{4}+\sqrt{\left(\frac{c_s^{   -2}-1}{4}\right)^2+c_s^{   -2}p^2}   \\
 \ \ \ \ \ \ \ \ \ \ \ -p\ \ \ \ \ \ \ \ \ \ \ \ \ \ \ \ \ \ \ \ \ \ \ \ \  , & p
\end{array} \right),
\la{general}
\end{eqnarray}
while we single out the first component $c_0\,\beta_0(l)$,
whose derivative simply writes \eq \b'_{0}(l) =
\left(1-e^{2l}\right)^{\f{1- 1/{c_s^2}}2}
%\quad(\equiv c_0\ b'_0(l))
\la{first}. \eqx In \eqref{general}, $\,_2F_1(a,b;c,z)$ is the
usual hypergeometric function while $G^{2,\,0}_{2,\,2}
\left(e^{2l}\Big| \begin{array}{cc}
a\ \ , &b  \\
c \  \ , &d
\end{array} \right)$ denotes a Meijer  function \cite{grad}.

The function $\b_{p}^{(1)}$ ($resp.\ \b_{p}^{(2)}$) is the regular
($resp.$ irregular)  solution\footnote{For $p=0,$ Eq.
\eqref{coeffs} is only first-order for $\beta_0'(l)$ and thus
introduces only one arbitrary coefficient $c_0.$}  at $e^l=0$ of
\eqref{coeffs}. Hence, any arbitrary combination of these two
independent functions is a solution of \eqref{coeffs}. The
boundary conditions will define the specific linear combinations to be
chosen for the general solution which may be  obtained
from the Green function of the problem. For illustration, in the
special case $p=0,$ one writes
\begin{eqnarray}
\b'_{0}(l) = \int_{-\infty}^{+\infty}\ \Theta(-l)\
\left(1-e^{2(l-\hat{l})}\right)^{\frac{1-c_s^{-2}}{2}}
F_{0}(\hat{l})\ d\hat{l}, \la{sources}
\end{eqnarray}
where  $F_0(l)$ describes  a distribution of sources in
temperature convoluted with  the Green function, which in this
case is just $\Theta(-l)\,\b'_{0}(l) $ from \eqref{first}. A
straightforward but more tedious expression can be written for all
values of $p$ but is skipped here for brevity. A specific
realization for the sake of our physical
 problem will be discussed in the application section.

Inserting the general solution (\ref{solchi},\ref{general}) for
the KK potential in Eq. (\ref{dentropydphi}) for the azimuthal
entropy distribution, one finds
\begin{eqnarray}
&&\frac{dS}{d\vp}(\vp)=\frac{s T_0}{T}\left\{\,\beta'_0(l)
+\sum_{p=1}^\infty
\cos(2p\vp)\left[(2p)^2\b_{p}(l)-\b'_{p}(l)\right]
\right\}\nonumber \\
&&v_2(T)=\ \frac{4\b_1(l)-\b_1'(l)} {2 \b'_{0}(l) }\equiv
\rho\left[\ \frac{4\b^{(1)}_{1}(l)-\b'^{(1)}_{1}(l)} {2\b'_0 (l)}
+ \lam\
\frac{4\b^{(2)}_{1}(l)-\b'^{(2)}_{1}(l)}{2\b'_{0}(l)}\right]\ ,
\label{solv}
\end{eqnarray}
where we denote $\frac{c_{1}^{(1)}}{c_{0}}\equiv \rho,$ and
$\frac{c_{1}^{(2)}}{c_{1}^{(1)}}\equiv \lam,$ and
\begin{equation}
\ve(T)=\f{2\b_0'(l)\left[2\b_1(l)+\b'_1(l)\right]+\sum_{p=1}^\infty
\left[4p(p+1) \b_{p}(l)\b_{(p + 1)}(l)+2p\b_{p}(l)\b'_{p+1}(l)-
2(p + 1)\b_{p+1}(l)\b'_{p}(l)- \b'_{p}(l)\b'_{p+1}(l)\right]
}{2\b_0'^2(l)+\sum_{p=1}^\infty
\left[\left(2p\b_{p}(l)\right)^2+\b_{p}'^2(l)\right]}
\label{eccenexg}
\end{equation}
for the eccentricity. It is useful for further use to note that,
in the ``elliptic approximation'' $i.e.$ when one stops the
Fourier expansion \eqref{solchi} at $p= 1,$ the eccentricity  can
be expressed using the same functions and parameters with $v_2$ as
in Eq.\eqref{solv}, namely
\begin{equation}
\ve(T)=\frac{2\beta_0'(l)\left[2\b_1(l)+\b'_1(l)\right]}
{2\beta_0'^2(l)+\left[4\b_1^2(l)+\b_1'^{2}(l)\right]}\equiv
{\bf \rho} \ \left\{ \frac{2\beta_0'\left[2\left(\beta_1^{(1)}+
\lambda\beta_1^{(2)}\right)+\beta_1^{(1)'}+
\lambda\beta_1^{(2)'}\right]}
{2\beta_0^{'2}+{\bf \rho^2}\ \left[4\left(\beta_1^{(1)}+
\lambda\beta_1^{(2)}\right)^2+\left(\beta_1^{(1)'}+
\lambda\beta_1^{(2)'}\right)^2\right]}\right\}
\ .
\label{eps}
\end{equation}
Thanks to the analytic solutions we obtained, all the expressions
contain an explicit dependence in temperature. One should only
specify which temperature is physically relevant, $e.g.,$ $T_I$ for
the initial spatial eccentricity and some freeze-out temperature
$T_f$ for the observed $v_2.$ Using our formulae, one may
discuss the dynamical hydrodynamical process through the
temperature dependence of both the spatial and momentum average
anisotropy of the lump of quark-gluon plasma. For this sake, we
note an interesting  parameter-independent relation between the
spatial eccentricity at any temperature $T_I$ and $v_2$ at any
temperature $T_f$, namely
\begin{equation}
\f{v_2(T_f)}{\ve(T_I)} = \ \f{\b_1(T_f)\b'(T_I)}{\b_1(T_I)\b'(T_f)}
\times \f  {1-\f{\b'_1(T_f)}{2\b_1(T_f)}}{1+\f{\b'_1(T_I)}{2\b_1(T_I)}}
\times \left({\scriptstyle {\f12}}+{\scriptstyle {\f12}}\ \sqrt{1-2{ \ve^2(T_I)}\
\f{1+\f{\b'^2_1(T_I)}{4\b_1^2(T_I)}}
{\left(1+\f{\b'_1(T_I)}{2\b_1(T_I)}\right)^2}}\right)^{-1}\! .
\la{vepsFI}
\end{equation}

A general physical comment is in order about the parameters $\lam$
and $\rho$ defining the  relevant solutions in the ``elliptic
approximation''. Using a source of given temperature, the
parameter $\lam,$ which corresponds to the relative strength of
the two independent solutions of the second-order differential
equation \eqref{coeffs}, will specify the  initial condition of
the elliptic flow. The parameter $\rho,$ which is geometrical in
nature since it gives the relative strength of the elliptic
harmonic in \eqref{solchi}, will be related to the initial
centrality of the reaction. For general initial conditions,
the more general Green function formalism, $cf.$  \eqref{sources},
has to be used.

\section{Exact elliptic flow: applications}

Taking into account the  linear equations for the potential and
entropy distributions, $cf.$ (\ref{KK2},\ref{entropydistrib}), the
determination of the elliptic flow boils down to defining properly
the boundary conditions, $i.e.$  the sources of the hydrodynamic
expansion, which are given functions of temperature and azimuth.
In the following we will assume that the source is simply given by
a delta-function at the initial  temperature $T_I$ of the process
and a given initial eccentricity profile. We fix it by the
condition that $v_2(T_I)=0$ while $\ve(T_I)$ is maximal. Note
that the solution satisfies the constraint  $T_I \lesssim T_s,$
$i.e.$ the fluid is always supersonic.

In the ``elliptic approximation'' for which the Fourier expansion
of the potential \eqref{solchi} is limited to the two first
orders, the observables $v_2$ \eqref{solv} and $\ve$ \eqref{eps}
depend only on two relevant parameters, namely
$\rho=c_{1}^{(1)}/c_{0},$  obtained from the Fourier expansion
\eqref{solchi} and $\lam=c_{1}^{(2)}/c_{1}^{(1)},$ that is the
coefficient ratio between the regular and irregular solutions
\eqref{general} of the potential equations
(\ref{KK2},\ref{coeffs}).

\begin{figure}[ht]
\begin{center}
\mbox{\epsfig{figure=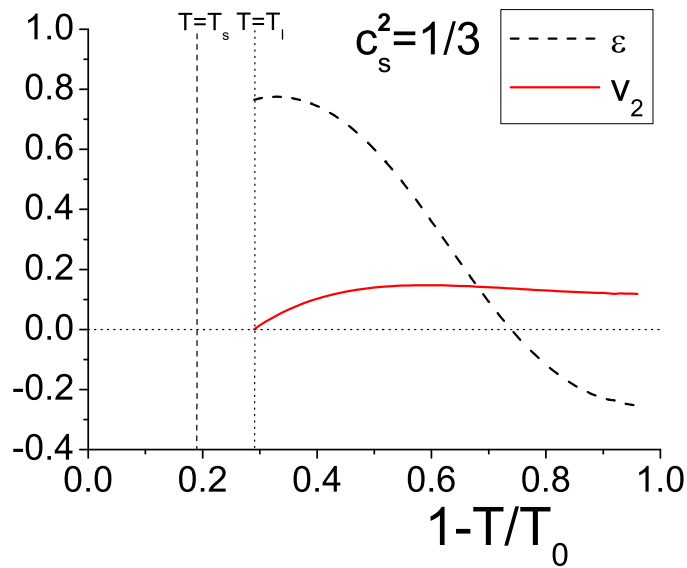,width=9cm,angle=0}} \caption{(Color
online) {\it Compared temperature dependence for the momentum
$v_2(T)$ and the spatial $\ve(T)$ anisotropies.} The curves
correspond to the initial temperature source at $T=T_I$ (see
text). The dashed line is for the supersonic lower bound  $T_s.$
The geometrical anisotropy parameter (see text) is $\rho=0.8$ and
the speed of sound
 is the reference one $c_s=1/\sqrt{3}.$}
\label{1}
\end{center}
\end{figure}

\vspace{.3cm}
\noindent{\it Determination of $\lam$}. From the previous
discussion,  $\lam$ is chosen in such a way that $v_2(T_I)=0,$
where $\ve(T_I)$ is maximal. As an illustration of the discussion,
the temperature dependence of both $v_2$ and $\ve$ that we obtain
with our definition of the initial condition, is displayed in
Fig.\ref{1}, for a given value of the geometrical anisotropy
parameter $\rho=0.8.$ The value of $T_I$ is lower but nearby the
speed-of-sound lower limit of temperature $T_s.$

\begin{figure}[ht]
\begin{center}
\mbox{\epsfig{figure=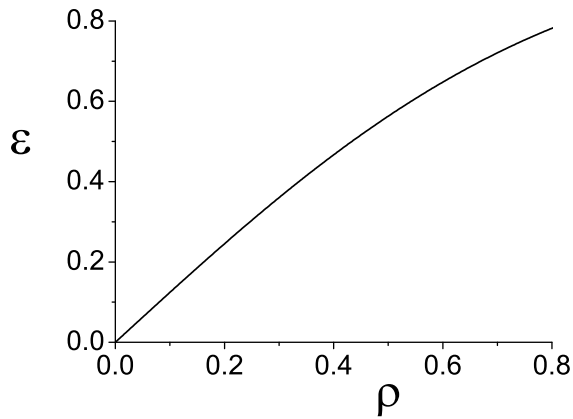,width=7.5cm,angle=0}} \caption{{\it
$\ve$ as a function of the geometrical anisotropy parameter
$\rho.$} The observed dependence qualitatively reproduces
simulations of $\ve$ as a function of centrality $c \sim
N_{part}/N_{max}$ (see text). In this figure we used $c_s^2=1/3$
and $\rho_{max}\sim 0.8$.} \label{2}
\end{center}
\end{figure}

\noindent {\it Determination of $\rho$}. The determination of the
geometrical parameter $\rho,$ the first anisotropy coefficient of
the potential $\chi,$ see \eqref{solchi}, is governed by the
centrality. In Fig.\ref{2} we display  $\ve(\rho)$
which shows a quasi-linear behavior.
This is in good agreement with the observed feature of the
experimentally reconstructed eccentricity with an observed
proportionality relation with the centrality $c \sim
N_{part}/N_{max},$ where $N_{part}$ is the number of participant
nucleons. Indeed, one expects a simple relation between  $\rho$
and  $c$ , up to a rescaling $\rho/\rho_{max}\sim
1\!-\!N_{part}/N_{max}$. With such a choice and using
Eq.\eqref{solv}, one finds a simple proportionality rule of $v_2$
with centrality, namely
 \eq
  v_2=\rho_{max}(1-c)\left[\
   \frac{4\b_{1}^{(1)}(l)-\b'^{(1)}_{1} (l)}
{2\beta'_0(l)} + \lam\ \frac{4\b_{1}^{(2)}(l)-
\b'^{(2)}_{1}(l)}{2\beta'_0(l)}\right]\ ,
  \la{distribN}
  \eqx
which is also expected from hydrodynamical simulations \cite{45}.
Note that, in this framework, $\rho_{max}$  is indeed independent
of the evolving temperature ratio $T/T_0$, but it may depend on
the initial conditions such as the type of heavy-ion reaction and
the initial c.o.m. energy (or $T_0$). The $T/T_0$
dependence, given in \eqref{distribN} by the function within
brackets,  is uniquely defined from \eqref{solv}. In our
calculations,  the linearity of the formula \eqref{solv} for $v_2$
in terms of the normalized second Fourier coefficient $\rho$ is a
direct consequence of the equation \eqref{KK2} for the KK
potential which, together with the azimuthal entropy distribution,
is diagonalized by the Fourier expansion \eqref{solchi}. It is
clear that the formulation of the initial eccentricity profile
depends on the initial conditions, and we take the curve in
Fig.\ref{2} as an example.

Fig.\ref{1} is interesting also from the point of view of the
dynamics of elliptic flow. Indeed, it is known from hydrodynamic
models  \cite{45} that the {\it momentum anisotropy}, represented
in our ``quasi-stationary'' approximation by the
temperature-dependent  $v_2(T),$ rapidly increases as a function
of proper time, and thus with decreasing temperature, to reach its
observed value. It is therefore confirmed to be a good indicator
of the early stage of the hydrodynamic expansion. On the same
footing, the {\it spatial} anisotropy, represented by $\ve(T),$
decreases as the system expands, even reaching negative values,
$i.e.$ changing the sign of the spatial anisotropy. We observe, in
Fig.\ref{1}, that the transversally isentropic flow follows the
same qualitative path as a function of temperature cooling. It is
also interesting to note that the final value of $v_2$ (and thus
the value of $v_2/\ve,$ where $\ve \equiv \ve(T_I)$ is the initial
eccentricity) is reached rather early and rather
 independently of the choice of the initial temperature for the transverse flow.

\begin{figure}[ht]
\begin{center}
\mbox{\epsfig{figure=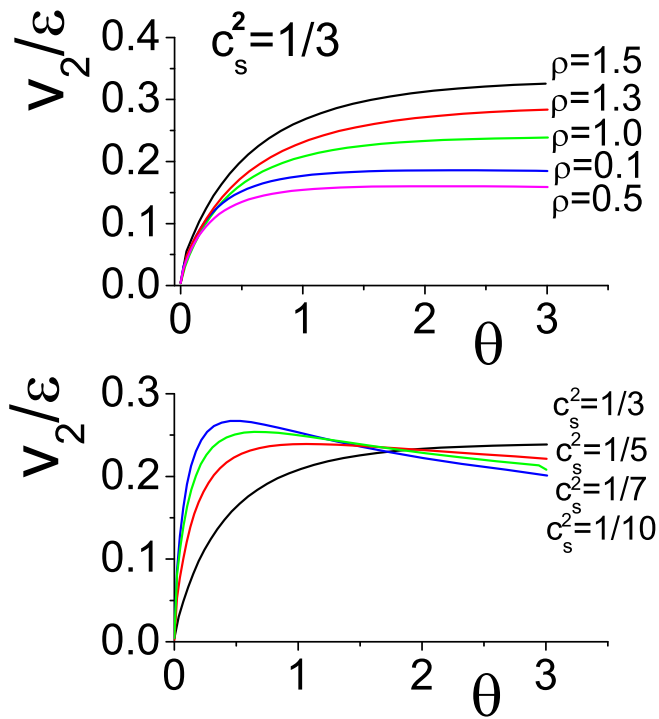,width=8.8cm,angle=0}}
\caption{(Color online)
 {\it
  $v_2/\ve$ as a function of the ``rescaled  time'' $\theta.$
  } Up:
  dependence on the centrality $via$ the  $\rho$-parameter at fixed $c_s=1/\sqrt 3;$
  Down: dependence on the EoS $via \ c_s^2 = (1/3,1/5,1/7,1/10)$ at fixed eccentricity $\ve=0.8$.
  For this sake, one
  is led to choose, respectively,  $\rho=(0.8, 1, 2, 4),$ by tuning the values of $\rho_{max}.$
  The ``reduced time'' is defined by Eq.\eqref{time}.}
\label{3}
\end{center}
\end{figure}

In order  to restore  the time variable through its dependence on
the temperature, we shall make use of  a convenient rescaling of
the temperature equivalent to the expansion time, similar to the
one proposed in \cite{bhalerao}, where the ratio
$v_2(\tau\!-\!\tau_0)/\ve(\tau_0)$ with initial time $\tau_0$ is
displayed for different values of impact-parameter and various
values of the speed-of-sound $c_s.$ One makes the rescaling substitution
 \eq
 \tau \ \to \ \f {c_s}{\bR}\ (\tau-\tau_0)\quad ;\quad
 \f1{\bR}=\sqrt{\f 1{\cor{x_1^2}}+\f 1{\cor{x_2^2}}}\ .
 \la{rescaling}
 \eqx
where $\bR$ gives an appropriate average estimate of the expanding
size of the plasma. In our temperature-dependent scheme, we define
an analogous rescaling using the thermodynamical relation
\eqref{temperature} by choosing a ``rescaled time'' variable
defined in terms of the temperature as \eq \theta \equiv \f
{c_s}{\bR} \ \left\{\ \left(\f {T_0}T\right)^{c_s^{-2}}-
 \left(\f {T_0}{T_I}\right)^{c_s^{-2}}\right\}\ .
 \la{time}
 \eqx
In Fig.\ref{3} we show the theoretical results for $v_2/\ve$
 as a function on the
rescaled variable $\theta$ for the solution we considered. Let us
comment both sides of the figure.
In the upper graph, the figure displays the dependence on
centrality $via$ $\rho =\rho_{max} (1\!-\!c).$  The value of
$\rho_{max}$ has been chosen fixing $v_2(\tau-\tau_0)/\ve(\tau_0)$
to match with some realistic value (see, $e.g.$ \cite{Hirano,45}).
One observes the general trend of the $\theta$ evolution as a
function of increasing centrality (or decreasing $\rho$). This
trend, which has been empirically observed in hydrodynamic
simulations \cite{bhalerao} is here explained by the nonlinear
$\ve$-dependent correction to $v_2/\ve$ (see \eqref{vepsFI}). It
is easy to realize that the remnant $\ve$-dependence in
\eqref{vepsFI} is such that it increases for increasing $\ve$ (the
denominator is smaller) and thus it decreases with centrality.
Hence our scheme reproduces, at least qualitatively, a trend as a
function of impact-parameter observed in hydrodynamical
simulations.

Also, in the lower graph of Fig. \ref{3} we display the
speed-of-sound dependence of $v_2(\theta)/\ve$. Thanks to the
``time'' rescaling \eqref{rescaling}, it is possible to
superimpose the different curves, provided an adequate choice of
$\rho$ ensures the constant initial $\ve(T_I).$ As also seen
numerically, the analytical dependence over $c_s$ of our resulting
formula \eqref{vepsFI} gives a  decreasing  value of the ratio
$v_2(\theta)/\ve(T_I)$ with decreasing speed-of-sound. Here also
one finds the observed behavior \cite{bhalerao}. However, this
hierarchy is obtained at rather larger $\theta$ than observed in
\cite{bhalerao}. We will comment on that feature in the next
section.

\section{Conclusion, discussion and outlook}

Let us briefly summarize our results: using the conjecture of a
``quasi-stationary'' and transversally isentropic hydrodynamic
regime governing the transverse flow, and for a given EoS, we
arrive at a closed system of hydrodynamic equations which can be
solved by a hodograph transform $x_{1,2}\to T,\vp\ .$ Thanks to the
potential method \cite{KK} we can formulate the general solution
and give explicit analytic expressions for the hydrodynamic
features of the transverse flow. In an application to a source
with given temperature and constant effective speed-of-sound
$c_s$, we are able to give a complete analytic solution. The
applications to the determination of the features of the elliptic
flow are in good qualitative agreement with the observed (or
numerical) characteristics: the temperature dependence of the
spatial and velocity anisotropy (see Fig.\ref{1}), the linear
behavior of $v_2$ with centrality ($cf.$ Eq.\eqref{distribN}) for
a realistic initial eccentricity (see Fig.\ref{2}) and the
centrality and speed-of-sound dependence of the ratio $v_2/\ve $
(see Fig.\ref{3}).

Now, possessing an analytic solution, it is possible to go back to
the initial assumptions and discuss their range of validity. In
other terms we may address the question to which approximation can
we consider our closed system of transverse equations to be a good
approximation of the full hydrodynamic equations. To quantify this
approximation  a meaningful  comparison is to give estimates of
two quantities which are relevant for the discussion of the two
hypotheses, {\bf a)} of a ``transversally isentropic'' and
{\bf b)}  a ``quasi-stationary transverse'' flow.

In order to test our conjecture {\bf a)} and looking to
\eqref{transentropy}, we are led to consider the following ratio
of entropy flow gradients \eq \f{\partial_{x_\bot}(s
u_\bot)}{\partial_{\tau}(s u_0)} \sim \f{\partial_{T}(s u_\bot)}
{\partial_{T}(s u_0)} / \ \f{\partial_{T}
x_\bot(T)}{\partial_{T}\tau(T)}\equiv \f1{\cal V}\
\f{\partial_{T}(s u_\bot)} {\partial_{T}(s u_0)}\ , \la{ratio1}
\eqx where ${\cal V}\equiv\f{\partial_{T}
x_\bot(T)}{\partial_{T}\tau(T)}$ is the average,
temperature-dependent, expansion rate. Indeed, this ratio governs
the effect of the time-gradient compared with a typical transverse
one. Note, however that the {\it overall} transverse entropy
 gradient is zero, by virtue of \eqref{transentropy}.

In \eqref{ratio1}, we have replaced the kinematical variables by
their temperature-dependent averages defined by our solution. The
approximation range of  a ``transversally isentropic'' is thus
related to the value of the (analytically known) expression
\eqref{ratio1} to be larger than 1 in a significant range of
``reduced time''. In the upper graph of Fig.\ref{4} one sees that
the transverse over longitudinal entropy gradient becomes indeed
significantly larger than 1 for high enough speed of sound. For
small speed of sound this requires longer ``reduced time''. This
could explain the features of Fig.\ref{3}, low, with a
``retarded'' ordering w.r.t. \cite{bhalerao}.

In order to test the consistency of the ``quasi-stationary''
approach {\bf b)}, in the lower graph of Fig.\ref{4} we present
the expansion rate itself ${\cal V}\equiv\f{\partial_{T}
x_\bot(T)}{\partial_{T}\tau(T)}$, where the functions
$\tau(T),x_\bot(T)$ are analytically obtained from their
definition within our temperature-dependent scheme, namely from
\eqref{temperature} and \eqref{alpha} respectively. Note that this
rate is also appearing in the denominator of \eqref{ratio1}, which
shows that the two hypotheses of  a ``transversally isentropic''
and ``quasi-stationary transverse'' flow are indeed connected,
since a slow motion gives rise to a high transverse over time
typical entropy gradient.
\begin{figure}[ht]
\begin{center}
\mbox{\epsfig{figure=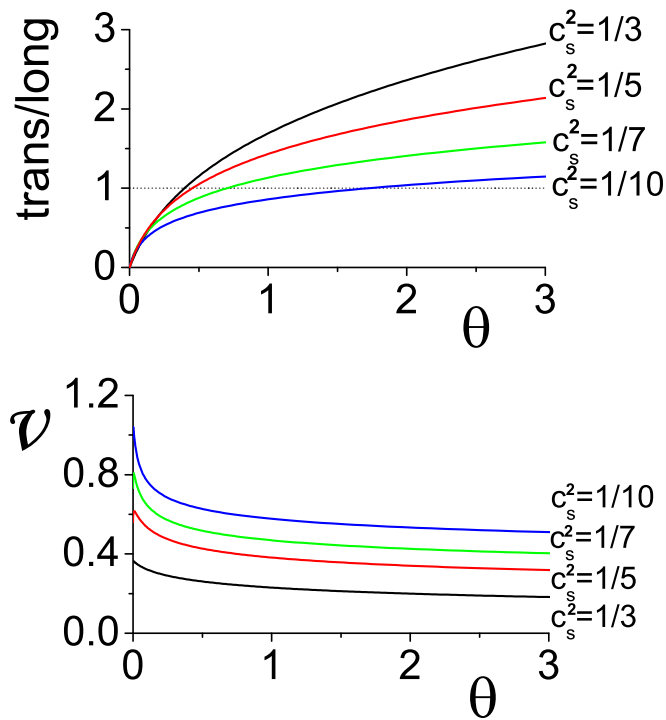,width=8.8cm,angle=0}}
\caption{(Color online) {\it Comparison of entropy and kinematic
gradients}. The analytically known quantities $\partial_{x_\bot}(s
u_\bot)/\partial_{\tau}(s u_0)$ and  ${\cal V}\equiv
\partial_{T}(x_\bot)/\partial_{T}\tau)$
(see text) are plotted as a function of the reduced time $\theta.$
} \la{4}
\end{center}
\end{figure}
From Fig.\ref{4} we see that both the transversally isentropic and
quasi-stationary hypotheses are consistent at not too short
``reduced times'' and not too small speed-of-sound. Thus, these
hypotheses give a qualitative analytic understanding of the
transverse flow. Our qualitative picture seems consistent.
However, the time gradient is not negligible w.r.t. the transverse
derivative, indicating, at least within the initial conditions we
choose, that a quantitative agreement could be more difficult to
be obtained.

Another topic is the range of validity of our approximation in
transverse space. Indeed, due to \eqref{sound}, the
quasi-stationary approach is only valid in the supersonic
dilatation regime, which requires a large enough transverse
velocity $v_\bot \ge c_s.$ This could limit the range of validity
of the hydrodynamical flow which has been observed only at small
transverse momentum. It could also compromise the dominance of the
longitudinal Bjorken  flow determining the thermodynamical
relations \eqref{temperature}. We think that this limitation,
which should be taken into account for a quantitative study
difficult to perform analytically, will not endanger the
qualitative but explicit solution we found. The study of the
implications for the transverse momentum dependence of the
elliptic flow deserves $per\ se$ a study which goes beyond the
scope of the present work, where no mass relation between fluid
velocity and transverse momentum has been introduced.

As an outlook,  it will be interesting to develop the study of
hydrodynamical mechanisms generating the elliptic flow by the
investigation of other phenomenological aspects, such as the
abovementioned $p_\bot$ dependence, the effect of a weak
viscosity, etc.... In order to reach more quantitative features,
it will be useful to refine the definition of the initial
conditions. In fact, it could be worthwhile to typically  defining
$a\ priori$ the dependence of $\ve$ as a function of $\rho$ or
centrality, and finding  the corresponding initial
conditions\footnote{We thank C.Gombeaud for this suggestion.} by
inverting $e.g.$ \eqref{sources}. In particular, to examine
whether they could identify more definitely the hydrodynamical
mechanisms. On a more theoretical ground, the existence of rather
simple mechanisms may facilitate the search for a relation to the
fundamental gauge field theory, and in particular the
gauge/gravity dual approach of the elliptic flow.

All in all, our results suggest  an analytic approach to the
transverse motion of the fluid, which can clarify the behavior of
the elliptic flow  obtained from the data or numerical
simulations. This is related to an approximate
``quasi-stationary'' and ``transversally isentropic'' property, of
the transverse flow  for
which the time dependence of the system comes mainly through the
temperature.

\section*{Acknowledgements}
We thank Jean-Yves Ollitrault for his suggestions and remarks and acknowledge fruitful
discussions with Guillaume Beuf, Andrzej Bialas, Cl\'ement Gombaud and  Wojciech Florkowski.
 We thank I.~M.~Khalatnikov and A.~Y~.~Kamenshchik for useful communication. One of us (E.N.S.)
 thanks the IPhT (Saclay) for hospitality during the achievement of this work.

\appendix
\section{Quasi-stationary transverse flow of a perfect fluid}\
\label{appstatflow}

In this section we derive the basic equations determining the
quasi-stationary transverse flow of a perfect fluid.
 The energy-momentum tensor
of a perfect fluid  is
\begin{equation}
 T^{\mu\nu}= ( e +p)u^{\mu}u^{\nu} - p \eta^{\mu\nu},  \label{Tmn}
\end{equation}
 where $ e $ is the energy density, $p$ is the
pressure and $u^{\mu}$ ($\mu =\{0,1,2,3\}$) is the  4-velocity in
the Minkowski metric $\eta^{\mu\nu}$, with signature
$(1,-1,-1,-1)$. It obeys the equations
\begin{equation}
\d_\mu T^{\,\mu}_\nu=0\ \ \ \ \Rightarrow\ \ \ \
u_\nu\partial_\mu\left[( e +p)u^\mu\right]+( e
+p)u^\mu\partial_\mu u_\nu-\partial_\nu p=0\, ,
 \la{hydro}\end{equation}
 with $
 u_\nu u^\nu=1$,
and thus
\begin{equation}
 u_\nu \partial_\mu u^\nu=0.\label{upartu}
\end{equation}
From now on and for simplicity, $\partial_\mu$ denotes
$\partial_{x_\mu}$. Multiplying the equations of motion
(\ref{hydro}) by $u_\nu$, i.e projecting them on the direction of
the 4-velocity, and using (\ref{upartu}) we acquire:
\begin{equation}
\partial_\mu\left[( e +p)u^\mu\right]-u^\mu\partial_\mu
 p=0.
 \la{proju}\end{equation}
Finally, re-inserting (\ref{proju}) into (\ref{hydro}) we obtain:
\begin{equation}
( e +p)u^\mu\partial_\mu u_\nu=-u_\nu u^\mu\partial_\mu
p+\partial_\nu p.
 \la{proju2}\end{equation}
Relation (\ref{proju2}) holds for a general perfect fluid. For a
stationary flow it gives rise to Bernoulli equation \cite{Landau},
namely
\begin{equation}
Tu_0=  T_0. \label{bern2}
 \end{equation}

Let us now focus on the stationary transverse flow \cite{KK},
namely setting $u_3=0$, which is the regime of interest of the
present work. In this case the  equation of motion (\ref{proju2})
for $u_3$ gives $\partial_3 p=0$ and thus the various quantities
do not depend on the longitudinal coordinate $x_3$. Therefore, $ e
$, $p$ and the velocities are functions of $x_1$, $x_2$ only. The
equations of motion (\ref{proju2}) boil down to:
\begin{eqnarray}
 \partial_1\left[( e +p)u_1^2+p\right]+ \partial_2\left[( e +p)u_1u_2\right]&=&0
\nonumber \\
 \partial_1\left[( e +p)u_1u_2\right]+ \partial_2\left[( e +p)u_2^2+p\right]&=&0
\nonumber \\
  \partial_1\left[( e +p)u_1u_0\right]+ \partial_2\left[( e +p)u_2u_0\right]&=&0\,, \label{eqbasic0}
\end{eqnarray}
with $ u_0^2=1+u_1^2+u_2^2$. Equations (\ref{eqbasic0}) can be
expressed in terms of the temperature and the entropy density.
Considering vanishing chemical potential we have:
\begin{equation}
p+ e  = Ts\;;\;\; d e  = T ds\;;\;\; dp = s dT \la{therm}.
\end{equation}
Using relations (\ref{therm}), equations (\ref{eqbasic0}) become:
\begin{eqnarray}
 \partial_1\left[(Ts)u_1^2\right]+ s \partial_1T+\partial_2\left[(Ts)u_1u_2\right]&=&0
 \label{eqbasic1}\\
 \partial_1\left[(Ts)u_1u_2\right]+ \partial_2\left[(Ts)u_2^2\right]+ s \partial_2T&=&0\label{eqbasic2}
 \\
  \partial_1\left[(Ts)u_1u_0\right]+ \partial_2\left[(Ts)u_2u_0\right]&=&0. \label{eqbasic3}
\end{eqnarray}
Now, the Bernoulli equation (\ref{bern2}) allows to  transform
(\ref{eqbasic3}) to the entropy conservation in the transverse
plane:
  \begin{equation}
 \partial_1(s u_1)+ \partial_2(s u_2)=0. \label{eqfin2}
\end{equation}
Finally, equations (\ref{eqbasic1}) and (\ref{eqbasic2}), with the
use of (\ref{upartu}), (\ref{bern2}), (\ref{eqbasic3}),  give rise
to the same equation:
\begin{equation}
 \partial_1(T u_2)=\partial_2(T u_1).\label{eqfin1}
  \end{equation}

\end{document}